\documentclass[pre,twocolumn,showpacs,aps,epsfig,superscriptaddress]{revtex4}

\usepackage{epsfig}

\begin{document}

\newcommand{\be}{\begin{eqnarray}}
\newcommand{\ee}{\end{eqnarray}}

\title{Measurements of the bulk and interfacial velocity profiles in
oscillating Newtonian and Maxwellian fluids}


\author{M. Torralba}
\affiliation{Departament d'Estructura i Constituents de la Mat\`eria, Universitat de
Barcelona, Av.\ Diagonal 647, E-08028 Barcelona, Spain}

\author{J.R. Castrej\'on--Pita}
\author{A.A. Castrej\'on--Pita}
\author{G. Huelsz}
\author{J.A. del R\'{\i}o}
\affiliation{Centro de Investigaci\'on en Energ\'{\i}a, Universidad Nacional Aut\'onoma
de M\'exico, A.P. 34, 62580 Temixco (Morelos), M\'exico}

\author{J. Ort\'{\i}n}
 \affiliation{Departament d'Estructura i Constituents de la Mat\`eria, Universitat de Barcelona, Av.\ Diagonal 647,
E-08028 Barcelona, Spain}

\vspace{5mm}
\date{\today}


\begin{abstract}
We present the dynamic velocity profiles of a Newtonian fluid
(glycerol) and a viscoelastic Maxwell fluid (CPyCl/NaSal in water)
driven by an oscillating pressure gradient in a vertical
cylindrical pipe. The frequency range explored has been chosen to
include the first three resonance peaks of the dynamic
permeability of the viscoelastic fluid / pipe system. Three
different optical measurement techniques have been employed. Laser
Doppler Anemometry has been used to measure the magnitude of the
velocity at the centre of the liquid column.
Particle Image Velocimetry and Optical Deflectometry are used to
determine the velocity profiles at the bulk of the liquid column
and at the liquid--air interface respectively. The velocity
measurements in the bulk are in good agreement with the
theoretical predictions of a linear theory. The results, however,
show dramatic differences in the dynamic behaviour of Newtonian
and viscoelastic fluids, and demonstrate the importance of
resonance phenomena in viscoelastic fluid flows, biofluids in particular,
in confined geometries.
\end{abstract}



\maketitle

\section{Introduction}

Coupling between flow and liquid structure makes the dynamic
response of non--Newtonian (complex) fluids much richer than that
of Newtonian (simple) fluids \cite{Gelbart96,Larson99}. In
particular, depending on the relevant time scale of the flow,
viscoelastic fluids exhibit the dissipative behaviour of ordinary
viscous liquids and the elastic response of solids. Due to their
elastic properties, these fluids are potential candidates to
exhibit interesting resonance phenomena under different flow
conditions.

In this respect, the response of a viscoelastic fluid to an
oscillatory pressure gradient has been analysed theoretically in
some detail. The response, measured in terms of the velocity for a
given amplitude of the pressure gradient,
exceeds that of an ordinary fluid by several orders of magnitude
at a number of resonant frequencies. The remarkable enhancement in
the dynamic response of the viscoelastic fluid is attributed to a
resonant effect due to the elastic behaviour of the fluid and the
geometry of the container
\cite{LopezdeHaro94,LopezdeHaro96,delRio98,Tsiklauri01}.

This theoretical prediction has been recently confirmed by Laser
Doppler Anemometry (LDA) measurements of velocity at the centre of
a fluid column driven by an oscillating pressure gradient
\cite{Castrejon03}. The experiments show that a Newtonian fluid
exhibits a simple dissipative behaviour, while a Maxwell fluid
(the simplest viscoelastic fluid) exhibits the resonant behaviour
predicted by the linear theory at the expected driving
frequencies. While this analysis was performed in the central
point of the fluid column, an exploration of the whole velocity
field is necessary to ensure that the linear model captures the
main features of the flow.

Resonance effects can only be observed when the elastic behaviour
of the fluid is dominant. This is properly characterised by a
dimensionless number De$\gg 1$, where De (Deborah number) measures
the relative importance of the relaxation time of the fluid to the
typical time scale of the flow. Moreover, at the resonant driving
frequencies the system response is nearly purely elastic and the
dissipative effects are very small.

In the present paper we extend the previous experimental
investigation in several directions. In the first place we present
new LDA measurements for two Maxwell fluids of different
composition, and show that the linear theory correctly predicts
the location of the resonance frequencies in terms of material
parameters. In addition, measurements at different depths along
the column centre show the attenuation of the velocity as the
upper free interface is approached. The main result is the
determination of the velocity profiles in the radial direction of
the cylindrical tube, at several time intervals and driving
frequencies. We have measured the velocity profiles at two
different locations: the bulk of the liquid column, and the upper
liquid--air interface. Particle Image Velocimetry (PIV) has been
used in the former case, and an original technique based on
Optical Deflectometry (OD) in the latter. The PIV results compare
satisfactorily with the velocity profiles predicted by the linear
theory \cite{delRio-Castrejon03}. The OD results show the
attenuation of the velocity due to surface tension.

\section{Linear theory}

In this section we recall the theoretical expression of the velocity
field in the bulk of a viscoelastic fluid contained in a vertical
tube, and subjected to an oscillating pressure gradient
\cite{delRio-Castrejon03,Castrejon03}. The tube is supposed to be
infinite in the vertical direction. The derivation is based on a
linear approximation of the hydrodynamic equations and a linearized
constitutive equation for a Maxwell fluid
\cite{LopezdeHaro94,LopezdeHaro96,delRio98}.

Following Ref.\ \cite{Castrejon03}, the velocity field in Fourier's
frequency domain reads:
\be
V(r,\omega)=-\frac{(1-i\omega t_m)}{\eta \beta^2} \left( 1 -
\frac{J_0 (\beta r)}{J_0 (\beta a)} \right) \frac{dP}{dz}.
\label{eq:V}
\ee
In this expression $r$, $z$ are the radial and vertical cylindrical
coordinates in a reference system centered with the tube axis,
$\omega = 2\pi \nu$, where $\nu$ is the driving frequency, $t_m$ is the
relaxation time of the Maxwell fluid, $\eta$ its dynamic viscosity,
$\beta = \sqrt{ \rho \left[ (t_m \omega)^2 + i \omega t_m
\right]/(\eta t_m) }$, $a$ is the cylinder radius, $J_0$ is the
cylindrical Bessel function of zeroth order, and $P$ is the applied
pressure in Fourier's frequency domain.

\begin{figure}
\begin{center}
\centerline{\epsfig{file=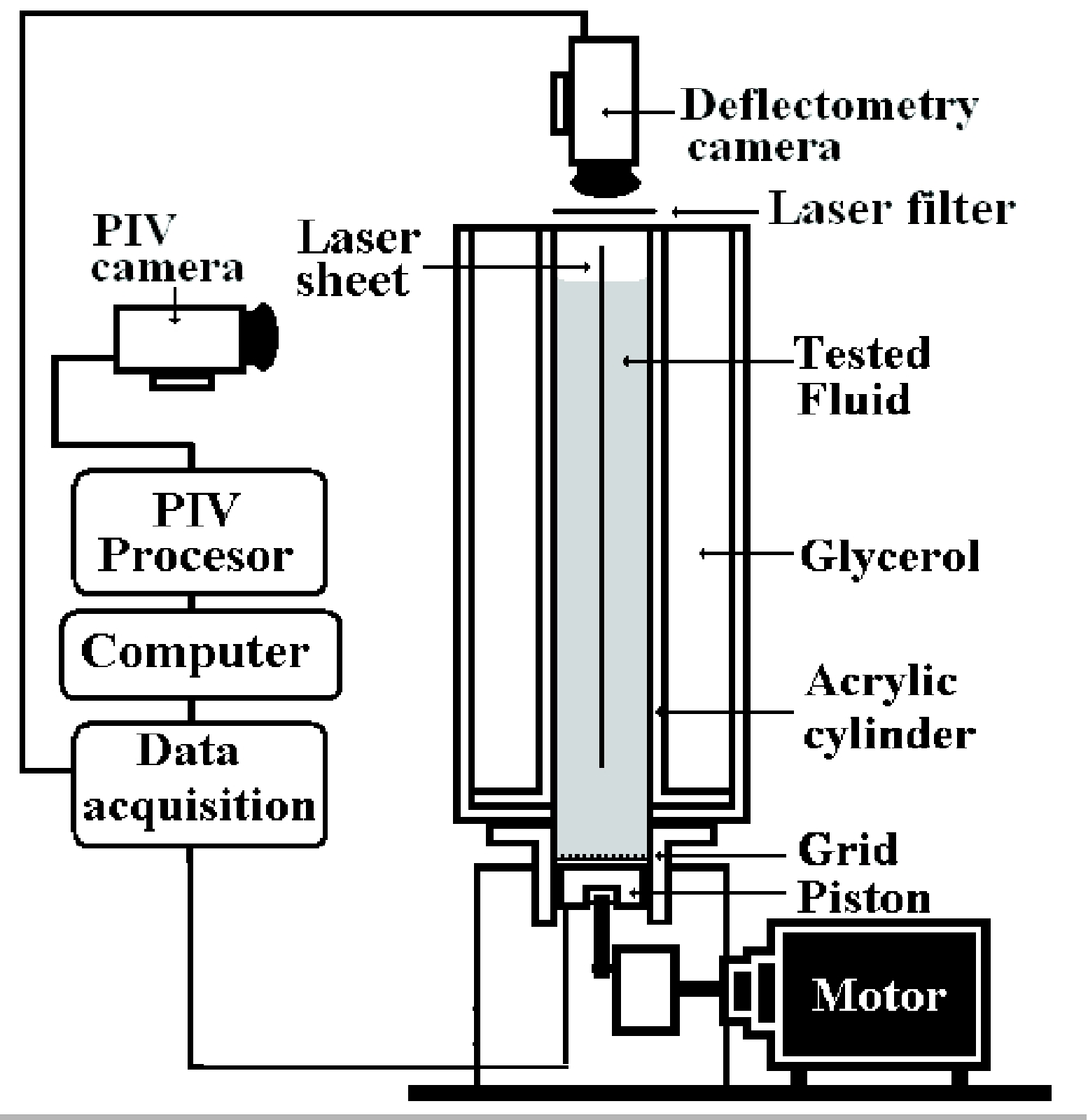,width=7cm}}
\vspace{3mm}
\caption{Schematic view of the experimental device, including the setups for PIV
and deflectometry measurements.}
\label{fig:setup}
\end{center}
\end{figure}

\begin{figure}
\begin{center}
\centerline{\epsfig{file=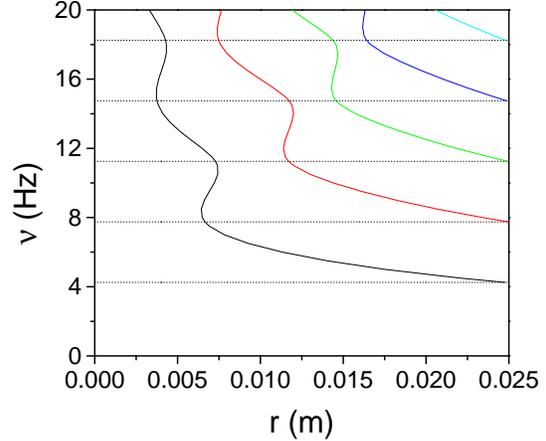,width=7cm}}
\vspace{3mm}
\caption{Diagram showing the location of the quiescent flow points
along the radial coordinate of the cylinder, $r$, as a function of
driving frequency, $\nu$, for a Maxwell fluid. The dashed
horizontal lines separate the frequency axis in intervals of
constant number of quiescent flow points. The parameters $\rho$,
$t_m$, and $\eta$ used to compute the diagram are the ones listed
in the text for the 60/100 CPyCl/NaSal solution, together with the
cylinder radius $a = 25$ mm.} \label{fig:stagnation-points}
\end{center}
\end{figure}

In our experiment, oscillations of the pressure gradient are induced by the
harmonic motion of a piston at the bottom end of the fluid column, with
amplitude $z_0$ (Figure \ref{fig:setup}).
Thus,
\be
\frac{dp(t)}{dz} = \rho z_0 \omega^2 \sin( \omega t).
\ee
The velocity profile in the radial direction of the tube, at
the driving frequency $\omega$, is then given by the real part of
the following expression:
\be
v(r,t)=-i \left( 1 - \frac{J_0
(\beta r)}{J_0 (\beta a)} \right) z_0 \omega e^{i \omega t}.
\label{eq:v-profile}
\ee
For a Newtonian fluid ($t_m = 0$) there is only one node of the
velocity profile, at $r = a$, which accounts for the non--slip
condition at the wall. Remarkably, if $t_m \neq 0$ (Maxwellian
fluid) the velocity profile may present several nodes. These nodes
define quiescent points of the flow. For given material and
geometrical parameters, the location of the nodes depends on the
driving frequency, as shown in Fig.\ \ref{fig:stagnation-points}.

As mentioned in the Introduction, the elastic properties of the
fluid dominate the dynamic response of the system at the resonant
frequencies. In Eq.\ (\ref{eq:v-profile}) this is manifest in
that, at the resonant frequencies, the oscillation in the centre
of the tube (maximum velocity) is in phase with the driving.

\section{Experimental device}

The experimental device is shown in Fig.\ \ref{fig:setup}. The
cylindrical container (inner radius $a = 25$ mm, length 500 mm) is
made of transparent acrylic. In order to avoid optical aberrations
this cylinder is placed inside a second recipient of square
section, made also of transparent acrylic, which is filled with
glycerol to match the refractive index of the acrylic walls. A
Teflon piston driven by a motor of variable frequency produces
harmonic oscillations of the pressure gradient in the liquid
column. Although the motor controller can operate in the frequency
range from $1.5$ to 200 Hz, our experiments have been performed in
the reduced frequency range from $1.5$ to 15 Hz. The piston
oscillation amplitude is set to $0.80 \pm 0.05$ mm to keep the
Reynolds number low, within the accessible frequency range, for
the two fluids studied. Additional details of this part of the
setup are described in Ref.\ \cite{Castrejon03}.

The viscoelastic fluid used in these experiments is an aqueous
solution of cetylpyridinium chloride and sodium salicylate
(CPyCl/NaSal). This surfactant solution is known to exhibit the
rheological behaviour of a linear Maxwell fluid in a range of
concentrations \cite{Hoffman91,Berret93}. In most experiments we
have used a 60/100 concentration, for which the dynamic viscosity
$\eta = 60$ Pa$\cdot$s, the density $\rho = 1050$ kg/m$^3$, and
the relaxation time $t_m = 1.9$ s \cite{MendezSanchez03}. In DLA
measurements we have also used a 40/40 concentration, for which
$\eta = 30$ Pa$\cdot$s, $\rho = 1005$ kg/m$^3$, and relaxation
time around $t_m = 1.25$ s. Our rheological characterisation of
this last fluid, however, suggests that it cannot be fully
described by a single relaxation time. As Newtonian fluid we have
used commercial glycerol, with nominal dynamic viscosity $\eta =
1$ Pa$\cdot$s and density $\rho = 1250$ kg/m$^3$. These values
have been determined at the working temperature of $(25 \pm
0.5)^\circ$C.

All sets of measurements performed on the same fluid have been
carried out in order of increasing frequency. The cylinder has
been emptied and refilled with fresh fluid every time that a
series of measurements at three different driving frequencies was
complete, i.e. approximately every day. Measurements in the bulk
have been carried out at two different heights, corresponding to
distances of 6 and 10 cm from the upper free interface.

The LDA technique used in the measurements presented in the next
Section has already been described in detail in Ref.\
\cite{Castrejon03}. We present now a brief description of the
other two techniques (PIV and deflectometry).

\subsection{Particle Image Velocimetry}

The PIV technique provides instant measures of the velocity maps
in a plane of the flow \cite{Adrian91}. To this purpose, the fluid
is seeded with small particles. The velocity maps are obtained by
a measure of the statistical correlation of the displacement of
the seeding particles in the fluid in a known time interval, in
this case the time between two consecutive laser pulses.

Our PIV system contains a two pulsed Nd-YAG lasers unit, that
includes an optical array to produce a laser light sheet in a
vertical plane of the acrylic cylinder (Fig. \ref{fig:setup}). A
high resolution camera (Kodak E1.0), perpendicular to the laser
light sheet, is used to record the digital images. The camera
records two consecutive frames, one corresponding to each laser
light pulse. The acquisition rate is limited by the camera to
three pairs of images every two seconds (1.5 Hz). A {\it Dantec
FlowMap 1100} processor takes care of the synchronization between
the laser pulses and the camera trigger. Post--processing of the
data, to determine velocity maps, is carried out by the {\it
Dantec FlowMap v5.1} software. {\it Dantec} 20--$\mu$m polyamid
spheres were used as seeding particles in the present experiments.
These particles are small enough to follow the flow with minimal
drag, but sufficiently large to scatter enough light to obtain
good particle images.

\subsection{Optical Deflectometry}

\begin{figure}
\begin{center}
\centerline{\epsfig{file=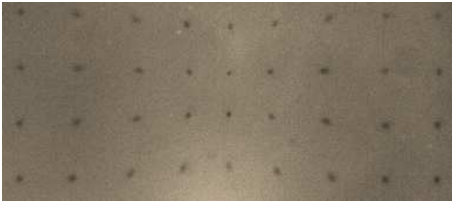,width=6cm}} \vspace{4mm}
\caption{A typical picture recorded in Optical Deflectometry
measurements. The image of a regular array of black dots (grid
spacing $d = 5.0$ mm) is distorted by the interface deformation.}
\label{fig:example-deflectometry}
\end{center}
\end{figure}

Deflectometry is a technique used to study relatively small
deformations in transparent fluid surfaces \cite{Fermigier92}. In
our case the liquid column serves as a variable--thickness lens. A
thin transparent plastic sheet with a regular array (grid spacing
$d$) of black dots (0.5 mm diameter) is placed on top of the
Teflon piston, at the bottom of the acrylic cylinder (Fig.\
\ref{fig:setup}). The grid is imaged through the liquid column
using a CCD camera interfaced to a computer. Light passing through
the liquid column is refracted at the liquid--air interface,
distorting the image of the grid as the interface deforms. Figure
\ref{fig:example-deflectometry} presents an example of the
recorded images.


\begin{figure}
\begin{center}
\centerline{\epsfig{file=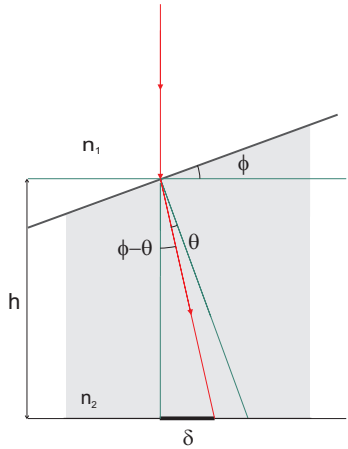,width=4.5cm}}
\vspace{3mm}
\caption{Sketch of the path followed by a light ray in the system.}
\label{fig:deflecto}
\end{center}
\end{figure}

The displacement of an imaged grid point, $\delta$, is related to
the local slope of the interface, $\phi$, in the following way
(Fig.\ \ref{fig:deflecto}). According to Snells' refraction law,
$n_1 \theta = n_2 \phi$. On the other hand, for a column height
$h$ the figure shows that $\tan(\phi - \theta) = \delta / h$.
Therefore, the relation between $\phi$ and $\delta$ reads:

\be
\phi = \frac{1}{1-n_2/n_1} \arctan \left( \frac{\delta}{h}
\right). \label{Eq:alpha-delta}
\ee

In our setup the height of the liquid column ($H = 240$ mm) is
much larger than the local height variations produced by the
deformation of the interface (of the order of mm), and we can
safely take $h \simeq H$ in Eq.\ (\ref{Eq:alpha-delta}).

OD is usually used in thin films. The deflection of the points of
the grid increases with the depth of the layer, making this
technique useless for the measurement of highly deformed
interfaces of thick layers. However, the small amplitude of the
perturbation in our experiments generates relatively smooth
profiles, and the condition of small deformations is satisfied in
a wide range of frequencies.

\begin{figure}
\begin{center}
\epsfig{width=6cm,file=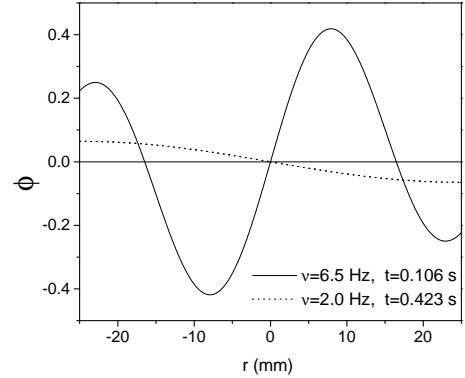}
\vspace{3mm}
\caption{Theoretical
deformation profiles of the Maxwell fluid driven at 2 Hz (dashed line)
and at $6.5$ Hz (solid line).} \label{Fig:theo-def-profiles}
\end{center}
\end{figure}

Although the deformations are small in this frequency range, the
whole deformation profile depends strongly on the frequency of
driving. This can be seen on the theoretical deformation profiles
of the viscoelastic fluid at 2 and $6.5$ Hz, displayed in Fig.\
\ref{Fig:theo-def-profiles}. The two profiles are very different.
While the deformation profile at 2 Hz is monotonic and very
smooth, the profile at $6.5$ Hz shows a more complex behaviour:
instead of a single central node, it presents two nodes which, in
the velocity profile, correspond to a maximum at the tube centre
and a minimum near the tube wall. The maximum deformation between
these two nodes corresponds to an inflection point of the velocity
profile, very close to the quiescent flow point.

In our setup the height of the liquid column is a function of the
radial coordinate $r$, and periodic in time. This height,
$h(r,t)$, is simply related to the local slope by
$\phi(r,t)=\partial h(r,t)/\partial r$. Hence:

\be
h(r,t) = h(r_0,t) + \int_{r_0}^r \phi(r,t) \, dr. \ee
It is useful to
choose for $r_0$ a position where the interface is motionless,
because the reference height $h(r_0,t)$ is then a constant that
can be taken equal to 0. We have chosen $r_0$ as the radial
coordinate of the quiescent flow point (in the bulk) closest to the
cylinder axis. Implicitly, we are assuming that the stagnation
points predicted by Eq.\ (\ref{eq:v-profile}) for the bulk do not
change their position as the interface is approached. This
assumption is confirmed by our PIV and OD results (see Sec.\
\ref{Sec:results}).


Finally, the velocity profile of the interface can be obtained
from the time derivative of the local height, $v(r,t) = \partial
h(r,t)/\partial t$.

In our experiments the grid spacings used are $d = 5.0$ and $5.5$
mm, and the refractive indices are $n = 1.33407(01)$ for the
CPyCl/NaSal solution (measured by Abb\'e refractometry), and $n =
1.473$ for glycerol (nominal).

\section{Results and discussion}\label{Sec:results}

\subsection{rms velocity at the cylinder axis}

The amplitude of the velocity field given by Eq.\
(\ref{eq:V}) presents resonance peaks at several resonance
frequencies. As mentioned in the Introduction, this phenomenon was
demonstrated experimentally and compared to the purely dissipative
behaviour of a Newtonian fluid in Ref.\ \cite{Castrejon03}. These
results showed that the linear theory gives a good prediction of
the resonance frequencies but overestimates the amplitude of the
resonance peaks.

We have repeated this same kind of measurements in the present
work, to check the dependence of the resonance frequencies on the
rheological properties of the Maxwell fluid. To this purpose the
driving frequency $\omega$ has been rescaled by a characteristic
time $\tau$ defined as \be \tau = 10^{2/5} t_m \sqrt{\alpha}, \ee
where $\alpha$ is the inverse of our Deborah number $De = t_m \eta
/ a^2 \rho$. As it was shown in Ref.\ \cite{delRio98}, the
location of the resonance peaks becomes universal (independent of
fluid parameters and system dimensions) if $\omega$ is made
dimensionless in the form $\omega \tau$.

\begin{figure}[tbp]
\begin{center}
\epsfig{width=7.5cm,file=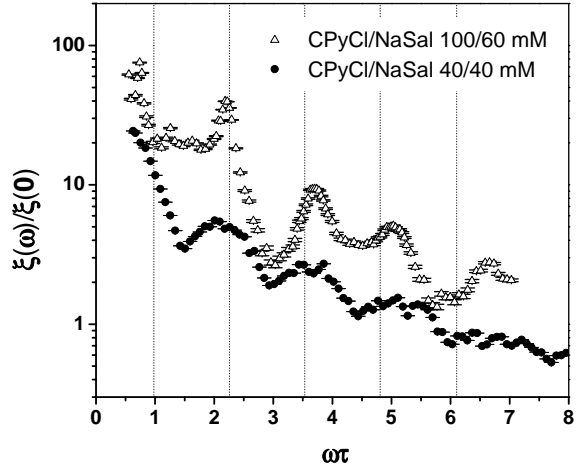}
\vspace{3mm}
\caption{Dimensionless response function at the centre of the
cylinder, as a function of dimensionless driving frequency, for
two concentrations of the CPyCl/NaSal solution. The vertical dashed lines
give the location of the resonance frequencies predicted by a linear theory.}
\label{Fig:LDA}
\end{center}
\end{figure}

Then, we rewrite Eq.\ (\ref{eq:V}) in the form
\be
V(r,\omega) = \xi(r,\omega) \frac{dP}{dz},
\ee
and plot the dimensionless response function $\xi(r,\omega) / \xi(r,0)$ as a function of the
dimensionless driving fequency $\omega \tau$. The results at the centre of the tube ($r=0$),
for the two different
concentrations of the CPyCl/NaSal solution used, are shown in Fig.\ \ref{Fig:LDA}.
Comparison with the linear theory is given by the vertical dashed lines, which give the
resonance frequencies corresponding to maxima of  $\xi(r=0,\omega)$. The Figure shows that
the rescaling of the frequency axis suggested by the linear theory leads to a satisfactory
reproducibility of the resonance frequencies independently of viscoelastic
fluid parameters.

The magnitude of the response function at the resonance peaks for
the surfactant solution of concentration 60/100 is considerably
lower in the present measurements than in the previous ones of
Ref.\ \cite{Castrejon03}. The reason is that the present
measurements have been carried out at 6 cm of the free air--liquid
interface, while the previous ones had been taken at 10 cm. The
different measured velocities provide a first evidence of the
damping influence of the free interface on the flow.

The results presented in the next two sections provide measurements
of the whole velocity profile, instead of measurements at a single point
in the flow \cite{Castrejon03}. The velocity profiles are obtained
at two different heights, in the bulk of the fluid, and at the fluid-air
interface.

\subsection{Velocity profiles in the bulk}

\begin{figure}
\begin{center}
\epsfig{width=6cm,file=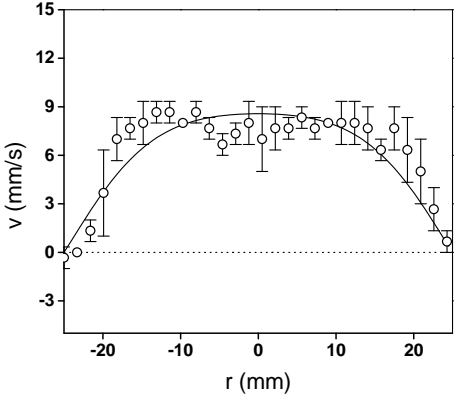}
\vspace{3mm}
\caption{Glycerol: PIV results at 2 Hz (dots),
measured at 6 cm from the upper free interface,
and the corresponding theoretical
prediction (solid line) at $t = 0.125$ s.}
\label{Fig:f1-2Hz}
\end{center}
\end{figure}

\begin{figure}
\begin{center}
\epsfig{width=6cm,file=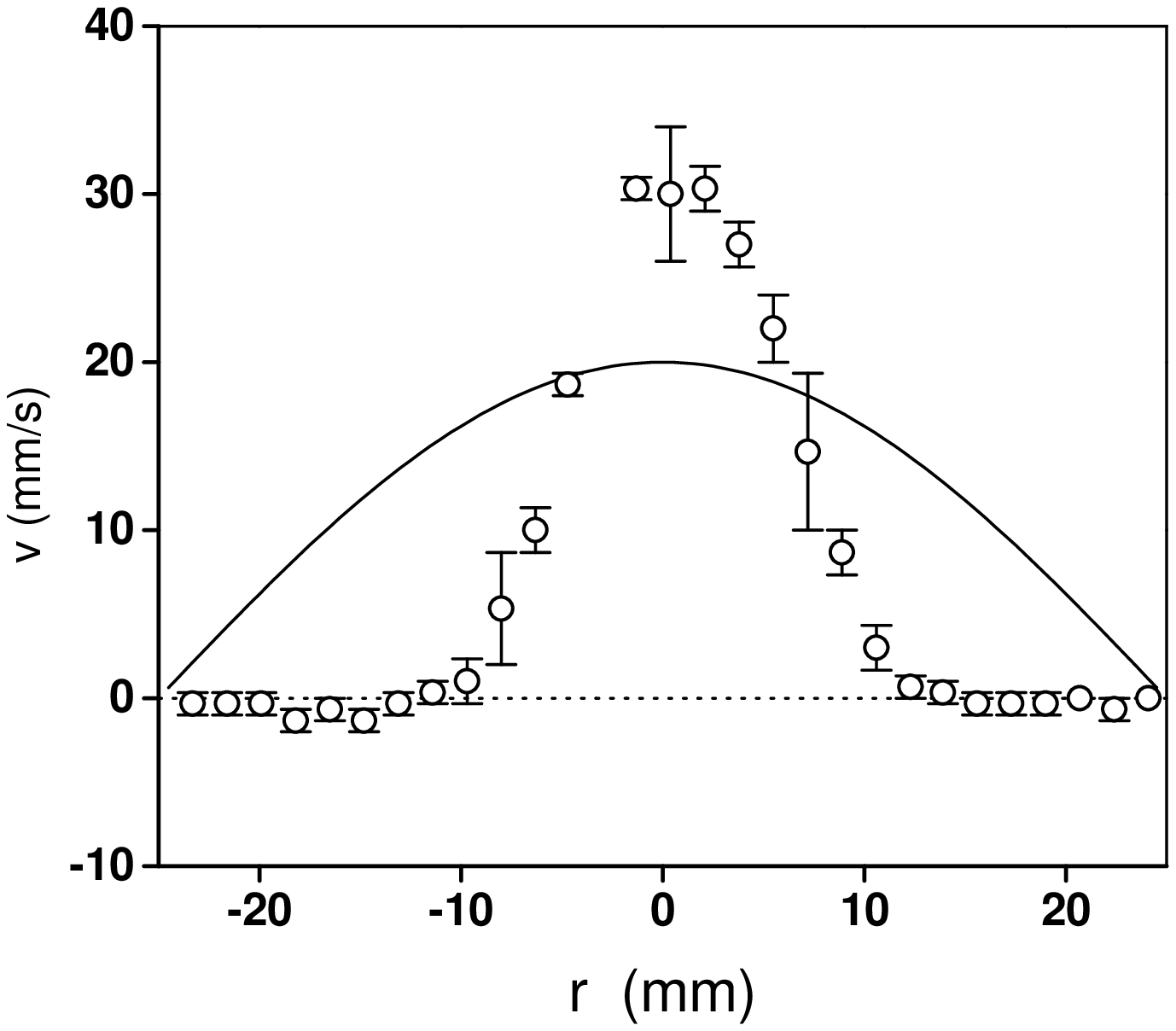}
\vspace{3mm}
\caption{60/100 CPyCl/NaSal solution: PIV results at 2 Hz (dots),
measured at 6 cm from the upper free interface,
and the corresponding theoretical
prediction (solid line) at $t = 0.375$ s.}
\label{Fig:f2-2Hz}
\end{center}
\end{figure}

\begin{figure}
\begin{center}
\epsfig{width=6cm,file=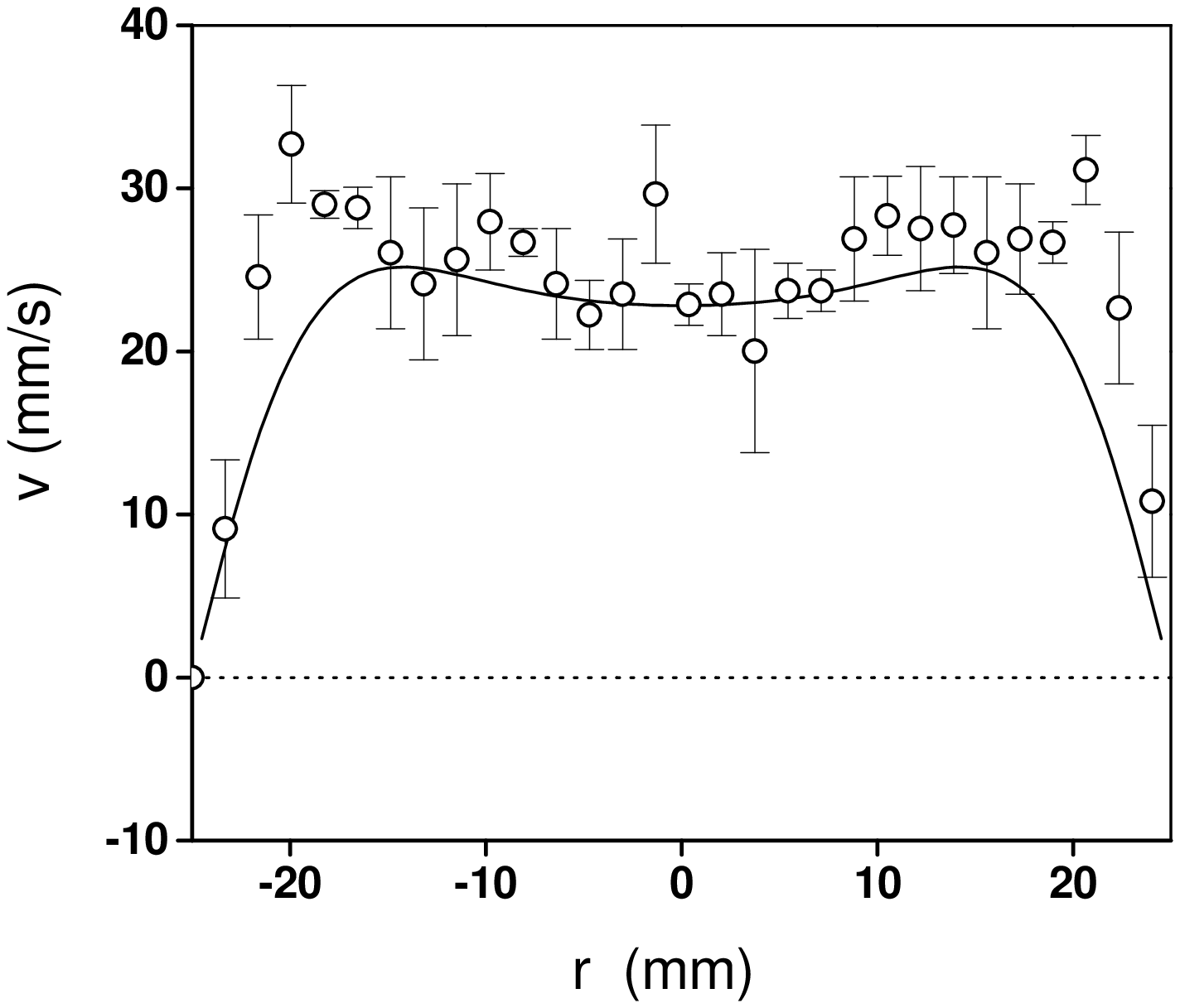}
\vspace{3mm}
\caption{Glycerol: PIV results at 6.5 Hz (dots),
measured at 6 cm from the upper free interface,
and the corresponding theoretical
prediction (solid line) at $t = 0.115$ s.}
\label{Fig:f1-6p5Hz}
\end{center}
\end{figure}

\begin{figure}
\begin{center}
\epsfig{width=6cm,file=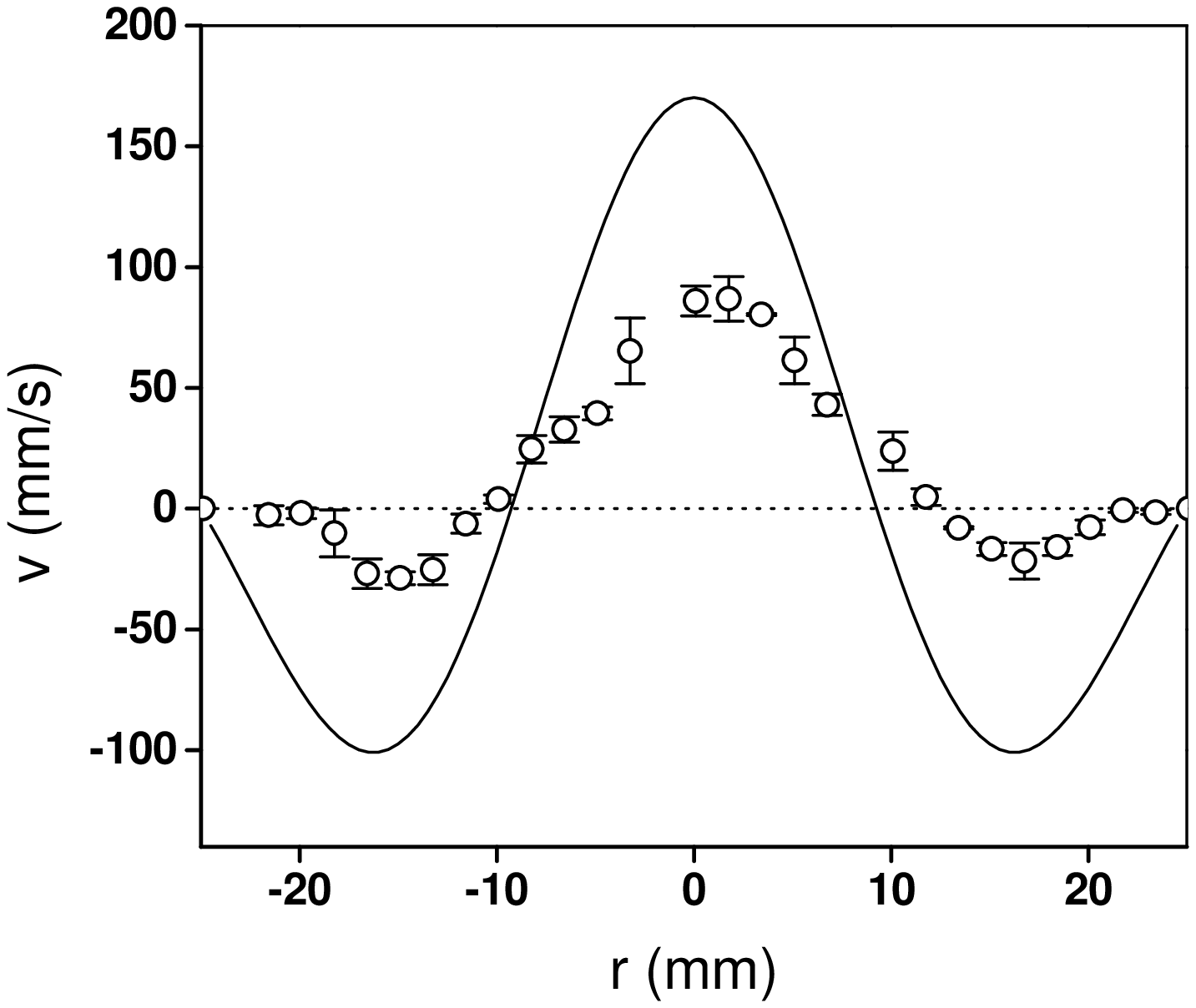}
\vspace{3mm}
\caption{60/100 CPyCl/NaSal solution: PIV results at 6.5 Hz (dots),
measured at 6 cm from the upper free interface,
and the corresponding theoretical
prediction (solid line) at $t = 0.038$ s.}
\label{Fig:f2-6p5Hz}
\end{center}
\end{figure}

\begin{figure}
\begin{center}
\epsfig{width=6cm,file=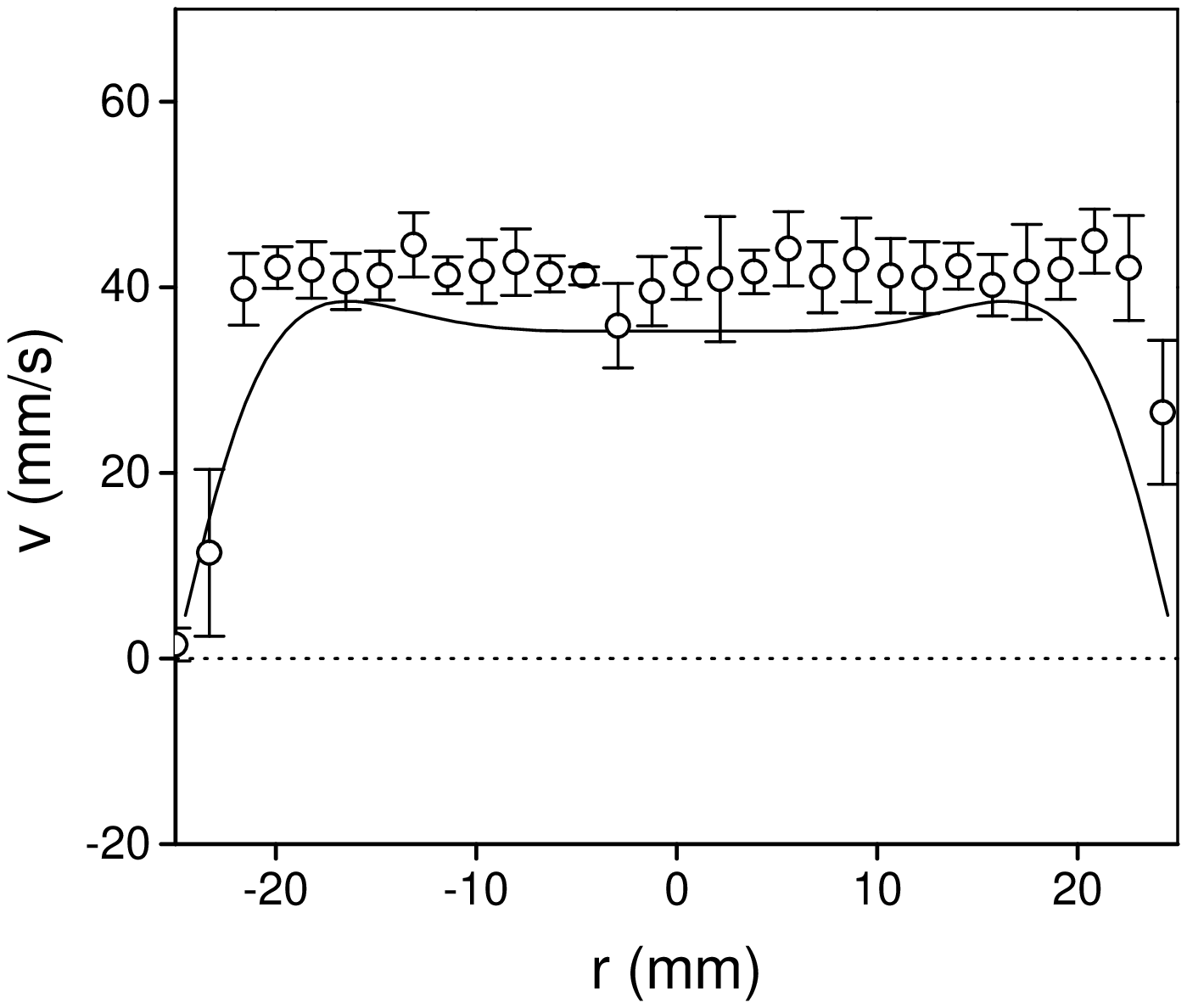}
\vspace{3mm}
\caption{Glycerol: PIV results at 10 Hz (dots),
measured at 6 cm from the upper free interface,
and the corresponding theoretical
prediction (solid line) at $t = 0.075$ s.}
\label{Fig:f1-10Hz}
\end{center}
\end{figure}

\begin{figure}
\begin{center}
\epsfig{width=6cm,file=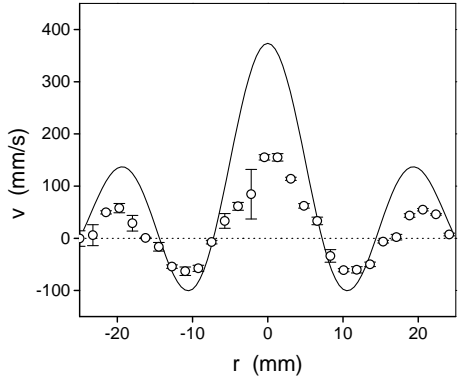}
\vspace{3mm}
\caption{60/100 CPyCl/NaSal solution: PIV results at 10 Hz (dots),
measured at 6 cm from the upper free interface,
and the corresponding theoretical
prediction (solid line) at $t = 0.025$ s.}
\label{Fig:f2-10Hz}
\end{center}
\end{figure}

The following set of figures presents the velocity profiles in the
bulk of the fluid column, determined by PIV measurements, together
with the theoretical profiles given by Eq.\ (\ref{eq:v-profile})
at coincident time--phases, for comparison. The profiles have been
determined at the driving frequencies of $2$ Hz (Figs.\
\ref{Fig:f1-2Hz} and \ref{Fig:f2-2Hz}), $6.5$ Hz (Figs.\
\ref{Fig:f1-6p5Hz} and \ref{Fig:f2-6p5Hz}), and $10$ Hz (Figs.\
\ref{Fig:f1-10Hz} and \ref{Fig:f2-10Hz}). These values coincide
with the first three resonance frequencies of our system for the
60/100 viscoelastic solution, as given by Eq.\
(\ref{eq:v-profile}). The time--phases have been selected by the
criterion that the velocity at the tube axis is a maximum.


The first observation to make is that the instantaneous velocity
profiles of the Maxwell fluid, driven at 2 Hz, present a single
defined sign of velocity (single direction of motion) along the
whole radius of the tube. As the driving frequency is increased to
6.5 and 10 Hz, however, the instantaneous profiles display a
progressively more complex structure, revealing the presence of
annular regions within the tube with alternating upward/downward
motion. Notice that this complexity is inherent to the
viscoelastic properties of the Maxwell fluid. For the Newtonian
fluid (glycerol) the instantaneous flow in the tube goes all in
the same direction for the three driving frequencies tested.

For the viscoelastic fluid the frontiers between
consecutive annular regions with alternating signs of the velocity
do not move. They correspond to the stagnation regions of the flow
which were already discussed in the context of the linear theory.
The PIV results show that the number of quiescent flow points along
the radial direction of the tube increases with the driving
frequency, in agreement with the theoretical prediction (Fig.\
\ref{fig:stagnation-points}).

\begin{figure}
\begin{center}
\epsfig{width=6cm,file=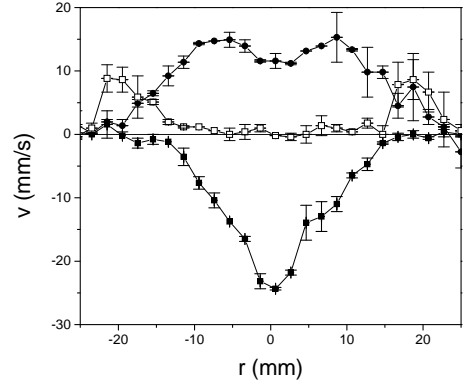}
\vspace{3mm}
\caption{60/100 CPyCl/NaSal solution: PIV results at 2 Hz,
measured at 6 cm from the upper free interface and
at different time phases within a driving period.}
\label{fig:vprofs-2Hz}
\end{center}
\end{figure}

Some of the measured profiles of the viscoelastic fluid show regions near the walls with vanishingly small
velocities and near zero velocity gradients, most noticeably the one at 2 Hz shown in Fig.\ \ref{Fig:f2-2Hz}.
These profiles are reminiscent of velocity profiles obtained for systems that display shear--banding. Indeed, the
CPyCl/NaSal solution is commonly known to show shear-banding \cite{Lourdes2}, but we have enough evidence to
discard this effect in our experiments. We did not observe an increase in turbidity in the region of the fluid
close to the walls, nor changes in the local intensity of the scattered light that could be attributed to
inhomogeneities. Furthermore, we monitored possible changes on the polarization state of the scattered light, an
indication of banding \cite{shear--banding}. No changes in the optical properties were observed for the amplitude
and frequency range used here. In addition, Fig.\ \ref{fig:vprofs-2Hz} demonstrates that the velocity near the
walls takes values distinctly different from zero at different phases within an oscillation period.

\begin{figure}
\begin{center}
\epsfig{width=6cm,file=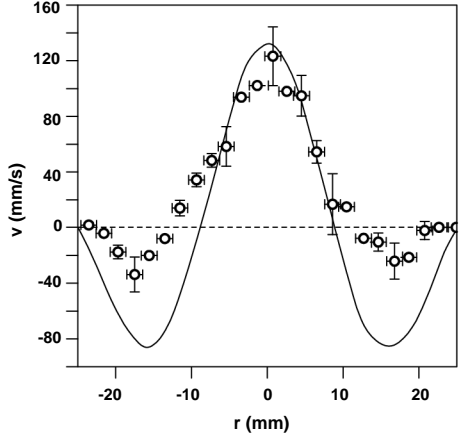}
\vspace{3mm}
\caption{60/100 CPyCl/NaSal solution: PIV results at 6.5 Hz (dots), obtained at 10 cm
from the upper free interface, and the corresponding theoretical
prediction (solid line) at $t = 0.057$ s.}
\label{fig:piv-10cm-6p5}
\end{center}
\end{figure}

\begin{figure}
\begin{center}
\epsfig{width=6cm,file=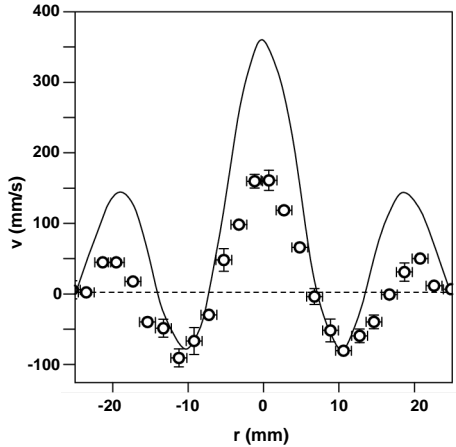}
\vspace{3mm}
\caption{60/100 CPyCl/NaSal solution: PIV results at 10 Hz (dots), obtained at 10 cm
from the upper free interface, and the corresponding theoretical
prediction (solid line) at $t = 0.022$ s.}
\label{fig:piv-10cm-10p5}
\end{center}
\end{figure}

Performing PIV measurements at two different heights of the liquid
column allows studying the influence of the upper free interface
on the velocity profiles. Thus, it is interesting to compare the
results presented above, which have been performed at 6 cm from
the upper interface, to the results shown in Figs.\
\ref{fig:piv-10cm-6p5} and \ref{fig:piv-10cm-10p5}, which have
been performed at 10 cm from the upper interface. The first
conclusion to draw from corresponding measurements at different
heights is that the location of the quiescent flow points is not
affected by the presence of the upper free interface. The second
conclusion is that the magnitude of the velocity profile is
smaller when the measurement is carried out closer to the free
interface. The damping effect of the free interface, disregarded
in the theory, originates from the air--liquid surface tension.
This observation applies equally to the two fluids investigated
(Newtonian and Maxwellian).

The experimental velocity profiles are in reasonable agreement
with the theoretical ones. The number of quiescent flow points is
the same, and their location very similar. As a general trend,
however, the theory overestimates the measured velocity.

A first explanation would be that the theory disregards nonlinearities,
and these tend to limit the largest velocity values.
Nonlinearities could arise from either the
hydrodynamic equations or the constitutive relation of the fluid.
In our case, however, since the small amplitude of the piston oscillations
ensures that Re$<$ 10$^{-4}$, the linearized momentum equation is
a very good approximation. On the other hand, taking into
account the cylindrical symmetry of the problem and assuming that the velocity
depends only on the radial coordinate (as our results confirm to a very good
approximation), it turns out that the first nonlinear correction to
the constitutive equation of the fluid cancels out exactly.

The disagreement between theory and experiment is probably due to shear--thinning of the viscoelastic fluid. As
shown in Ref.\ \cite{MendezSanchez03}, our fluid is properly described by a Maxwell model up to shear rates
$\gamma \simeq 0.6$ s$^{-1}$, and experiences shear--thinning beyond that value. A close inspection of
Figs.\ \ref{Fig:f2-2Hz}, \ref{Fig:f2-6p5Hz}, and \ref{Fig:f2-10Hz} reveals that the shear rate actually
experienced by the fluid in some phase intervals of the oscillation is larger \cite{Castrejon03}.
In these conditions the viscosity of the fluid decreases with shear. The theory predicts that the dynamic response
of the system at the resonant frequencies becomes smaller as the viscosity is reduced \cite{delRio98}, and thus would support
the view that the measured velocity profiles are systematically smaller than the theoretical ones for the
viscoelastic fluid (and not for the newtonian fluid) because of shear--thinning.

\subsection{Deflection of the air--liquid interface}

\begin{figure}
\begin{center}
\epsfig{width=7.5cm,file=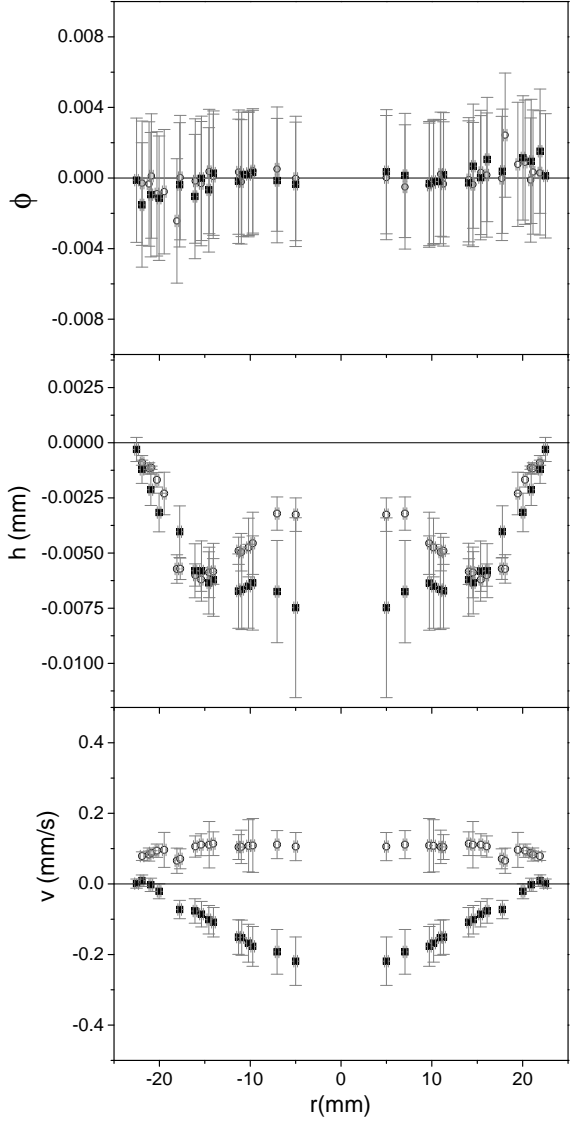}
\vspace{3mm}
\caption{Glycerol: deflectometry results at 2 Hz, measured at two different time--phases
within an oscillation.}
\label{Fig:f1-defl-2Hz}
\end{center}
\end{figure}

\begin{figure}
\begin{center}
\epsfig{width=7.5cm,file=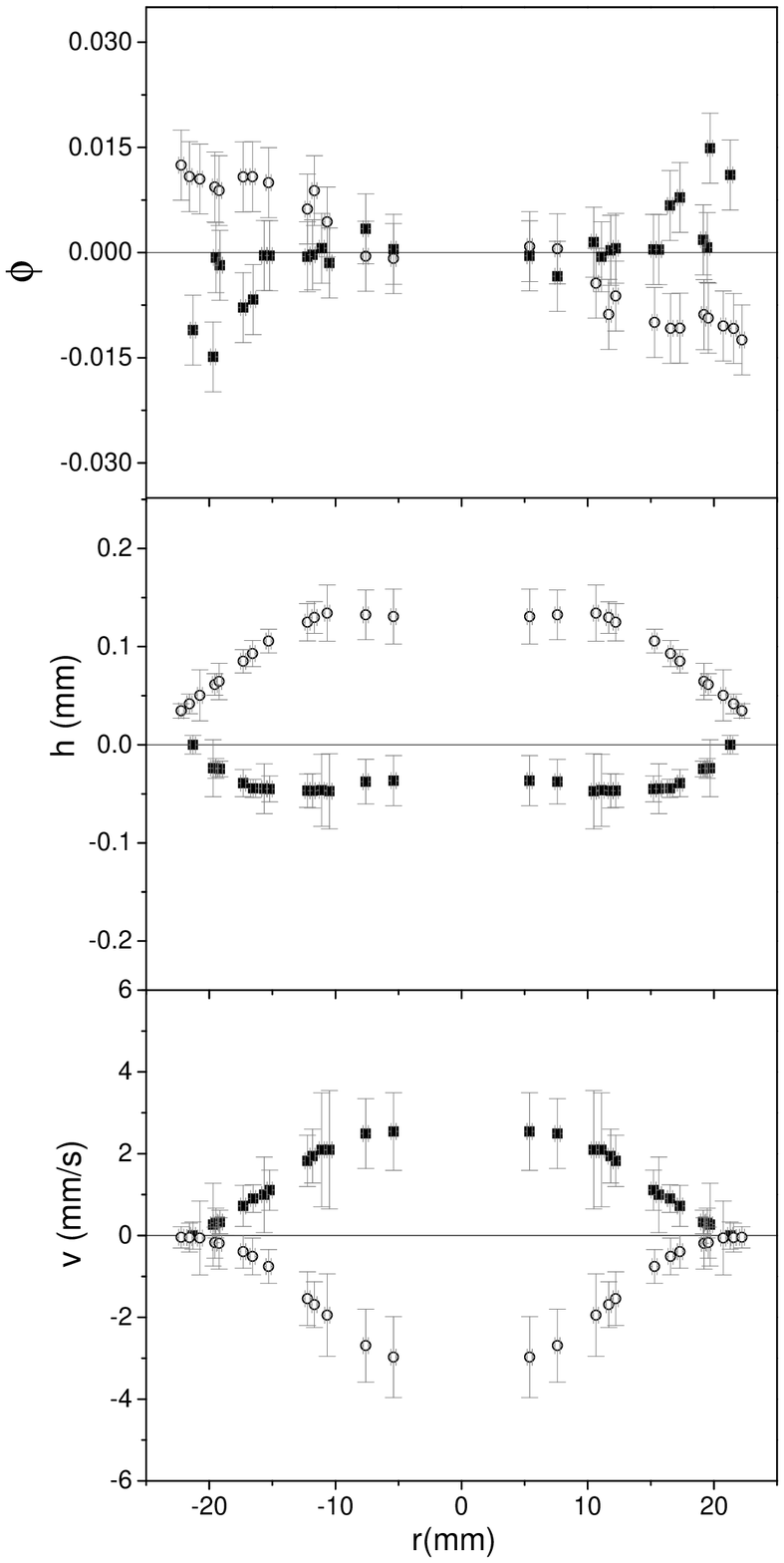}
\vspace{3mm}
\caption{60/100 CPyCl/NaSal solution: deflectometry results at 2 Hz, measured at two different
time--phases within an oscillation.}
\label{Fig:f2-defl-2Hz}
\end{center}
\end{figure}

\begin{figure}
\begin{center}
\epsfig{width=7.5cm,file=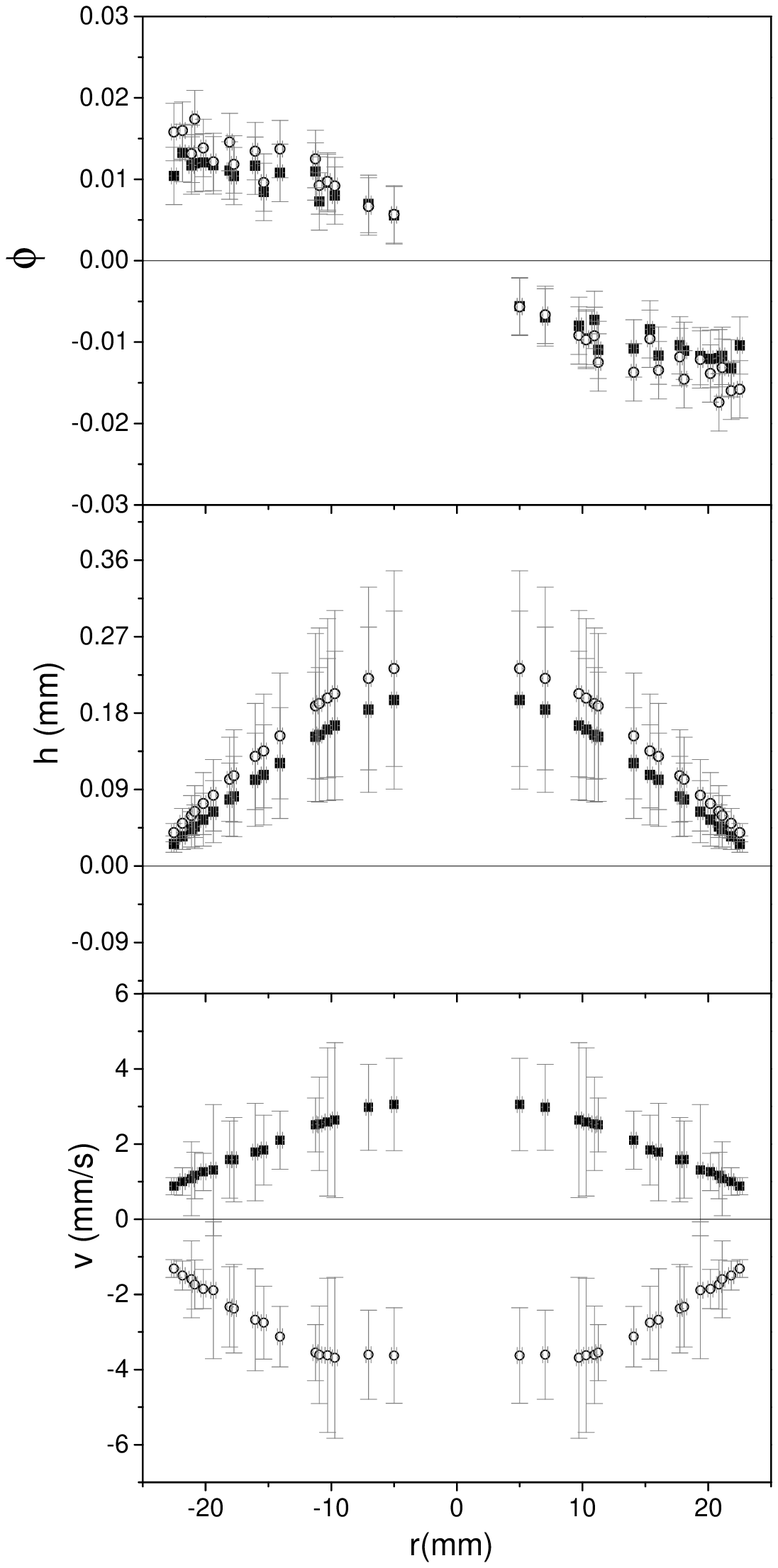}
\vspace{3mm}
\caption{Glycerol: deflectometry results at 6.5 Hz, measured at two different time--phases
within an oscillation.}
\label{Fig:f1-defl-6p5Hz}
\end{center}
\end{figure}

\begin{figure}
\begin{center}
\epsfig{width=7.5cm,file=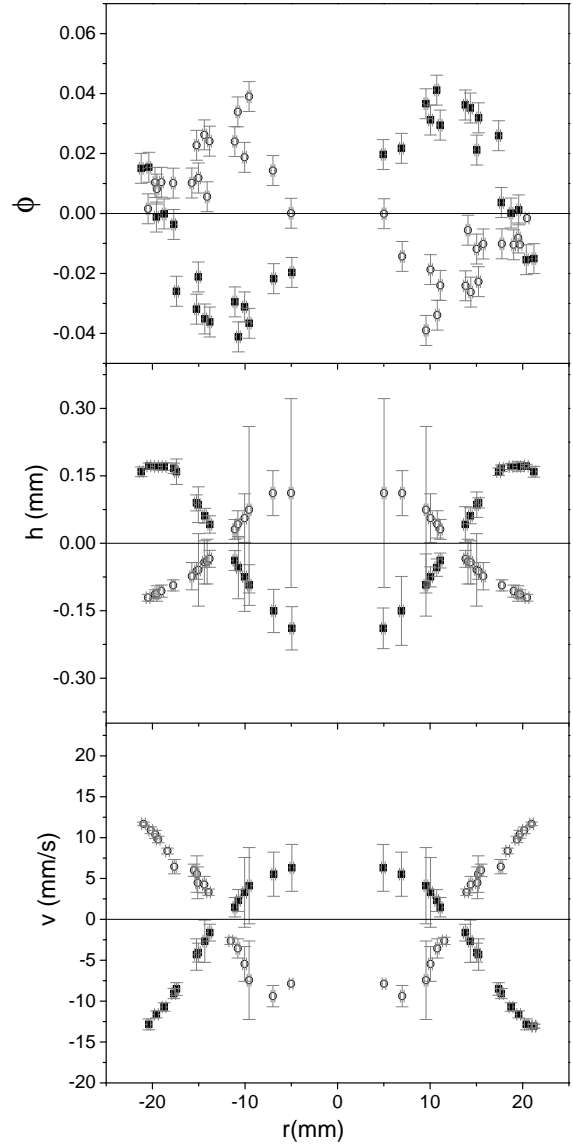}
\vspace{3mm}
\caption{60/100 CPyCl/NaSal solution: deflectometry results at 6.5 Hz, measured at
two different time--phases within an oscillation.}
\label{Fig:f2-defl-6p5Hz}
\end{center}
\end{figure}

The results obtained by Optical Deflectometry are presented on top
of Figs.\ \ref{Fig:f1-defl-2Hz} and \ref{Fig:f2-defl-2Hz} (2 Hz),
and Figs.\ \ref{Fig:f1-defl-6p5Hz} and \ref{Fig:f2-defl-6p5Hz}
($6.5$ Hz). There are no results at 10 Hz because, at this driving
frequency, the deformations of the free interface of the
Maxwellian fluid are so large that the technique is not
applicable. Each figure displays measurements at two different
time--phases.

The experimental deformation profiles of the interface at $2$ and
$6.5$ Hz for the viscoelastic fluid can be compared to the
theoretical ones (computed for the bulk) displayed in Fig.\
\ref{Fig:theo-def-profiles}. We observe that their shapes are
fully coincident for each of the two driving frequencies. This
shows that the quiescent flow points of the velocity profiles do not
change their position along the vertical direction, and thus their
location in the bulk can be used as reference to compute the
velocity profiles at the interface from the deformation profiles.

The velocity profiles are presented on the bottom of Figs.\
\ref{Fig:f1-defl-2Hz} and \ref{Fig:f2-defl-2Hz} (2 Hz), and Figs.\
\ref{Fig:f1-defl-6p5Hz} and \ref{Fig:f2-defl-6p5Hz} ($6.5$ Hz).
The magnitude of the interfacial velocities is in all cases much
lower than the one measured in the bulk of the fluid. This is
mostly due to the stabilizing role of surface tension, as
discussed above.

It is interesting to note that the velocity profiles of the interface depend also on the direction of motion of
the driving piston. The magnitude of the velocity is systematically lower for positive displacements of the piston
(liquid displacing air) than for negative ones (air displacing liquid). We attribute this asymmetry to the fact
that the large viscosity contrast across the interface stabilizes the displacement of a nearly inviscid fluid
(air) by a viscous liquid, and destabilizes the opposite displacement.


\section{Conclusions}

We have used three different optical techniques to characterize
the oscillating flow of a viscoelastic fluid (a solution of
CPyCl/NaSal in water) contained in a vertical cylinder and subjected
to an oscillating pressure gradient. A Newtonian fluid (glycerol) has
also been studied for the sake of comparison.

LDA measurements of the fluid velocity at the symmetry axis of the
cylinder as a function of driving frequency, for two different
compositions of the surfactant solution, have enabled us to show that
the frequencies of the resonance peaks can be predicted accurately in
terms of the fluid rheological properties by a simple linear theory. This
theory neglects inertial effects, and makes use of a linear
Maxwell model as constitutive relation for the viscoelastic fluid.

Systematic PIV measurements of the radial velocity field in the bulk
of the fluid have been performed at three different driving
frequencies, close to the first three resonance frequencies of the
viscoelastic system. While the velocity profile of the Newtonian
fluid along the radial direction does not change sign, this is not
the case for the Maxwellian fluid. The profiles measured at $6.5$ and
10 Hz present regions with alternating signs of the velocity,
separated by quiescent flow points. The number of quiescent flow points
increases with the driving frequency, revealing the increasing
complexity of the flow. Measurements within the fluid column at two
different heights show that these quiescent flow points do not shift as one
moves along the vertical direction, and that their radial location is
accurately reproduced by the linear theory.

The presence of a free interface at the top of the liquid column has
a damping effect on the velocity amplitude. This observation is
visible, both with LDA and PIV, when the results of measurements
carried out at two different heights within the liquid column are
compared.

Optical Deflectometry measurements of the free interface confirm that the velocity field is severely damped by the
surface tension of the air--liquid interface, compared to the velocity field within the bulk. Interestingly, the
deflectometry results show also that the oscillations of the velocity field at the interface are asymmetric, the
profiles corresponding to positive displacements of the piston having a slightly but systematically smaller
amplitude than those corresponding to negative displacements. We attribute this asymmetry to the fact that the
upward motion of the interface (liquid displacing air) is stabilized by the viscous pressure gradient in the
liquid. Recently it has been suggested that the viscous fingering interfacial instability of a viscoelastic fluid
in a Hele--Shaw cell will be enhanced by making the flow oscillate at the resonant frequencies \cite{Corvera04}.
Our deflectometry results show that the stabilizing role of the surface tension will reduce this effect to some
extent.

These results, as part of the study of resonance frequencies in viscoelastics, are
potentially relevant in biorheology to interpret the frequencies at which
biofluids are optimally pumped in living organisms \cite{Lambert04}.

\section{Acknowledgements}

We acknowledge A. Morozov (Universiteit Leiden) for illuminating discussions,
and J. Soriano (Universit\"{a}t
Bayreuth) and G. Hern\'{a}ndez (UNAM) for their technical
assistance. This research has received financial support through projects
BQU2003-05042-C02-02 and BFM2003-07749-C05-04 (MEC, Spain),
SGR-2000-00433 (DURSI, Generalitat de Catalunya), and CONACyT
38538 (Mexico). JO acknowledges additional support from the DURSI.
JRCP acknowledges the support by CONACyT, SEP, and the ORS Award
Scheme. AACP acknowledges the support given by the Dorothy Hodgkin
Award.



\begin{thebibliography}{100}

\bibitem{Gelbart96}
For an overview, see W.M. Gelbart and A. Ben--Shaul, J. Phys. Chem. {\bf 100},
13169 (1996).

\bibitem{Larson99}
R.G. Larson, {\it The Structure and Rheology of Complex Fluids} (Oxford
University, Oxford, 1999).



\bibitem{LopezdeHaro94}
M. L\'opez de Haro, J.A. del R\'{\i}o, and S. Whitaker, in {\it Lectures on
Thermodynamics and
Statistical Mechanics}, edited by M. Costas, R.F. Rodr\'{\i}guez, and A.L.
Benavides (World
Scientific, Singapore, 1994), pp. 54--70.

\bibitem{LopezdeHaro96}
M. L\'opez de Haro, J.A. del R\'{\i}o, and S. Whitaker, Transp. Porous Media
{\bf 25}, 167 (1996).

\bibitem{delRio98}
J.A. del R\'{\i}o, M. L\'opez de Haro, and S. Whitaker, Phys. Rev. E {\bf 58},
6323 (1998); {\bf 64}, 039901 (E) (2001).

\bibitem{Tsiklauri01}
D. Tsiklauri and I. Beresnev, Phys. Rev. E {\bf 63}, 046304 (2001).


\bibitem{Castrejon03}
J.R. Castrej\'on--Pita, J.A. del R\'{\i}o, A.A. Castrej\'on--Pita, and G.
Huelsz,
Phys. Rev. E {\bf 68}, 046301 (2003).

\bibitem{delRio-Castrejon03}
J.A. del R\'{\i}o and J.R. Castrej\'on--Pita, Rev. Mex. F\'{\i}s. {\bf 49}, 75
(2003).


\bibitem{Hoffman91}
R.H. Hoffman, Mol. Phys. {\bf 75}, 5 (1991).

\bibitem{Berret93}
J.F. Berret, J. Apell, and G. Porte, Langmuir {\bf 9}, 2851 (1993).

\bibitem{MendezSanchez03}
A.F. M\'endez--S\'anchez, M.R. L\'opez--Gonz\'alez, V.H. Rol\'on--Garrido, and J.
P\'erez--Gonz\'alez,
and L. de Vargas, Rheol. Acta {\bf 42}, 56 (2003).


\bibitem{Adrian91} R.J. Adrian, Annu. Rev. Fluid Mech. {\bf 23}, 261 (1991).

\bibitem{Fermigier92} M. Fermigier, L. Limat, J.E. Wesfreid, P. Boudinet, and C.
Quilliet,
J. Fluid Mech. {\bf 236}, 349 (1992).



\bibitem{Lourdes2} A.F. M\'endez-S\'anchez, J. P\'erez--Gonz\'alez, L. de Vargas, J.R. Castrej\'on--Pita,
A.A. Castrej\'on--Pita, and G. Huelsz, J. of Rheology {\bf 47}, 1455 (2003).

\bibitem{shear--banding} S. Lerouge and J.P. Decruppe, Langmuir {\bf 16}, 6464 (2000).



\bibitem{Corvera04} E. Corvera Poir\'e and J.A. del R\'{\i}o, J. Phys.: Condens. Matter
{\bf 16}, S2055 (2004).

\bibitem{Lambert04} A.A. Lambert, G. Ib\'a\~nez, S. Cuevas, and J.A. del
R\'{\i}o, Phys. Rev. E  {\bf 70}, 056302 (2004).


\end{thebibliography}
\end{document}